\documentstyle[12pt]{article}

\def\ref{\hangindent=0.7cm\hangafter=1}

\begin{document}

\centerline{\bf QUANTUM ALGEBRAIC SYMMETRIES}

\centerline{\bf   IN NUCLEI, MOLECULES,
AND ATOMIC CLUSTERS}
\bigskip
\centerline{ Dennis Bonatsos$^1$ and C. Daskaloyannis$^2$}

\centerline{$^1$ Institute of Nuclear Physics, N.C.S.R. ``Demokritos''}

\centerline{ GR-15310 Aghia Paraskevi, Attiki, Greece }

\centerline{$^2$ Department of Physics, Aristotle University of Thessaloniki}

\centerline{ GR-54006 Thessaloniki, Greece}
\vspace{0.3cm}

\centerline{\bf Abstract}
\medskip

Various applications of quantum algebraic techniques in nuclear 
structure physics and in molecular physics are briefly reviewed
and a recent application of these techniques to the structure of 
atomic clusters is discussed in more detail. 

\bigskip
{\bf 1. Introduction}
\medskip

Quantum algebras (also called quantum groups) \cite{Chari,BieL,KlimykS}
are deformed versions of the
usual Lie algebras, to which they reduce when the deformation parameter 
$q$ is set equal to unity. From the mathematical point of view they are 
Hopf algebras. Their use in physics became popular with the introduction 
\cite{Bie,Mac} of the $q$-deformed harmonic oscillator as a tool for 
providing a boson realization of the quantum algebra su$_q$(2), although 
similar mathematical structures had already been known \cite{ArikCoon}. 
Initially used for solving the quantum Yang--Baxter equation, quantum algebras
have subsequently found applications in several branches of physics, as, for 
example, in the description of spin chains, squeezed states \cite{McD}, 
hydrogen atom and hydrogen-like spectra \cite{KiblerN,ArikAH,DayiD},
rotational and vibrational nuclear and molecular spectra and in conformal 
field theories. By now much work has been done \cite{Cod,ArikCelik,ArikUM,MMZ}
on the $q$-deformed oscillator and its relativistic extensions 
\cite{MKas,ArikM}, and several kinds of generalized deformed oscillators 
\cite{Das,ArikDT,PLB307,PLB331} and generalized deformed su(2) algebras 
\cite{JPA26,Pan,CJP46,JMP38}  have been introduced.  

Here we shall confine ourselves to applications of quantum algebras in nuclear 
structure physics and in molecular physics. The purpose of this 
short review is to provide the  reader with references for further reading.
In addition a recent application of quantum algebraic techniques to 
the structure of atomic clusters will be discussed in more detail. 

\bigskip
{\bf 2.  The su$_q$(2) rotator model} 

\medskip
 The first application of quantum algebras in nuclear physics was the use 
of the deformed algebra su$_q$(2) for the description of the rotational 
spectra of deformed \cite{RRS,PLB251}
and superdeformed \cite{JPG17} nuclei. 
The Hamiltonian of the $q$-deformed rotator is proportional to the 
second order Casimir operator of the su$_q$(2) algebra. Its Taylor expansion 
contains powers of $J(J+1)$ (where $J$ is the angular momentum), being 
similar \cite{PLB251}
to the expansion provided by the Variable Moment of Inertia 
(VMI) model. Furthermore, the deformation 
parameter $\tau$ (with $q=e^{i\tau}$) has been found \cite{PLB251} 
to correspond to 
the softness parameter of the VMI model. Through a comparison 
of the su$_q$(2) model to the hybrid model
the deformation parameter $\tau$
has also been connected to the number of valence nucleon pairs \cite{MRR20}
and to the nuclear deformation $\beta$ \cite{MRR21}. 
Since $\tau$ is an indicator
of deviation from the pure su(2) symmetry, it is not surprising that 
$\tau$ decreases with increasing $\beta$ \cite{MRR21}. 
The su$_q$(2) model
has been recently extended to excited (beta and gamma) bands \cite{MDRRB}. 

B(E2) transition probabilities have also been described in this framework 
\cite{JPA3275}. 
In this case the $q$-deformed Clebsch--Gordan coefficients are used instead 
of the normal ones. (It should be noticed that the $q$-deformed angular 
momentum theory has already been much developed \cite{JPA3275}.) 
The model predicts 
an increase of the B(E2) values with angular momentum, while the rigid 
rotator model predicts saturation. Some experimental results supporting 
this prediction already exist \cite{JPA3275}. 
Similarly increasing B(E2) values are predicted by a modified version 
\cite{Muker} of the su(3) limit of the  Interacting Boson Model (IBM), by 
the su(3) limit of the sdg Interacting Boson Model \cite{LongJi}, 
by the Fermion Dynamical Symmetry Model (FDSM) \cite{PCWF}, 
as well as by the recent systematics of Zamfir and Casten \cite{ZC}. 

\bigskip
{\bf 3. Extensions of the su$_q$(2) model}
\medskip

The su$_q$(2) model has been successful in describing rotational nuclear 
spectra. For the description of vibrational and transitional nuclear 
spectra it has been found \cite{PRC50}
that $J(J+1)$ has to be replaced by $J(J+c)$.
The additional parameter $c$ allows for the description of nuclear 
anharmonicities in a way similar to that of the Interacting Boson Model
(IBM) and the Generalized Variable Moment of Inertia (GVMI) model
\cite{PRC29}. 
The use of $J(J+c)$ instead of $J(J+1)$ for vibrational and 
transitional nuclei is also supported by recent systematics 
\cite{DGGRR52}. 

Another generalization is based on the use of the deformed algebra 
su$_{\Phi}$(2) \cite{JPA26,Pan,CJP46,JMP38}, 
which is characterized by a structure function $\Phi$. 
The usual su(2) and su$_q$(2) algebras are obtained for specific choices 
of the structure function $\Phi$. The su$_{\Phi}$(2) algebra has been 
constructed so that its representation theory resembles as much as possible
the representation theory of the usual su(2) algebra. Using this technique 
one can construct, for example, a rotator having the same spectrum as the 
one given by the Holmberg--Lipas formula \cite{Holmb}. 
A two-parameter generalization of the su$_q$(2) model, labelled as 
su$_{qp}$(2), has also been successfully used for the description of 
superdeformed nuclear bands \cite{BMKibler}. 

\bigskip
{\bf 4.  Pairing correlations}
\medskip

It has been found \cite{JPAL101} 
that correlated fermion pairs coupled to zero 
angular 
momentum in a single-$j$ shell behave approximately as suitably defined 
$q$-deformed bosons. After performing the same boson mapping to a simple 
pairing Hamiltonian, one sees that the pairing energies are also correctly
reproduced up to the same order. The deformation parameter used ($\tau
=\ln q$) is found to be inversely proportional to the size of the shell, 
thus serving as a small parameter. 

The above mentioned system of correlated fermion pairs can be described 
{\sl exactly} by suitably defined generalized deformed bosons 
\cite{PLB278}. Then 
both the commutation relations are satisfied exactly and the pairing energies 
are reproduced exactly. The spectrum of the appropriate generalized 
deformed oscillator corresponds, up to first order perturbation theory,
to a harmonic oscillator with an $x^4$ perturbation. 

If one considers, in addition to the pairs coupled to zero angular momentum,
pairs coupled to non-zero angular momenta, one finds that an approximate 
description in terms of two suitably defined $q$-oscillators (one describing 
the $J=0$ pairs and the other corresponding to the $J\neq 0$ pairs) occurs
\cite{JPA1299}. 
The additional terms introduced by the deformation have been found
\cite{JPA1299}
to improve the description of the neutron pair separation energies 
of the Sn isotopes, with no extra parameter introduced. 

$q$-deformed versions of the pairing theory have also been given in 
\cite{ShaSha,AMen}. 

\bigskip
{\bf 5.  $q$-deformed versions of nuclear models}
\medskip

A $q$-deformed version of a two dimensional toy Interacting Boson Model (IBM) 
 with su$_q$(3) overall symmetry 
has been developed \cite{JPAL267,Niigata}, 
mainly for testing the ways in which 
spectra and transition probabilities are influenced by the $q$-deformation.
The question of possible complete breaking of the symmetry through 
$q$-deformation, i.e. the transition from the su$_q$(2) limiting symmetry
to the so$_q$(3) one has been examined \cite{Cseh25,Gupta20}. 
It has been found that 
such a transition is possible for complex values of the parameter $q$ 
\cite{Gupta20}. 
(For problems arising when using complex $q$ values see \cite{Falco}).
Complete breaking of the symmetry has also been considered in the framework 
of an su$_q$(2) model \cite{GCLGS}. 
It has also been found \cite{Mesa} that 
$q$-deformation leads (for specific range of values of the deformation 
parameter $\tau$, with $q=e^{i\tau}$) to a recovery of the u(3) symmetry 
in the framework of a simple Nilsson model including a spin-orbit term. 
Finally, the o$_q$(3) limit of the toy IBM model has been used for the 
description of $^{16}$O + $\alpha$ cluster states in $^{20}$Ne, with 
positive results \cite{Cseh19}. 

$q$-deformed versions of the o(6) and u(5) limits of the full IBM have been
discussed in \cite{WangYang,GuptaLudu,Pan50}. 
The $q$-deformation of the su(3) limit of IBM 
is a formidable problem, since the su$_q$(3) $\supset$ so$_q$(3) 
decomposition has for the moment been achieved only for completely symmetric 
su$_q$(3) irreducible representations \cite{Jeugt,RRTBI,RRITer}.

Furthermore a $q$-deformed version of the Moszkowski model 
has been developed \cite{MAP,FSMen}
and RPA modes have been studied \cite{JPA895} in it. A $q$-deformed 
Moszkowski model 
with cranking has also been studied \cite{JPG1209} 
in the mean-field approximation. 
It has been seen that the residual interaction simulated by the 
$q$-deformation is felt more strongly by states with large $J_z$. The 
possibility of using $q$-deformation in assimilating temperature effects is 
receiving attention, since it has also been found \cite{PLA192}
that this approach
can be used in describing thermal effects in the framework of a $q$-deformed 
Thouless model for supercoductivity. 

In addition, $q$-deformed versions of the Lipkin-Meshkov-Glick (LMG) 
model have been developed, both for the 2-level version of the model
in terms of an su$_q$(2) algebra \cite{AEGPL}, 
and for the 3-level version 
of the model in terms of an su$_q$(3) algebra \cite{BPP}. 

\bigskip
{\bf 6. Anisotropic quantum harmonic oscillator with rational ratios of 
frequencies} 
\medskip

The symmetries of the 3-dimensional anisotropic quantum harmonic oscillator 
with rational ratios of frequencies (RHO) are of high current interest in 
nuclear physics, since they are
the basic symmetries underlying the structure of superdeformed and 
hyperdeformed nuclei \cite{Nolan,Janssens}.   
The 2-dimensional RHO is also of interest, in connection with ``pancake'' 
nuclei \cite{Rae}, i.e. very oblate nuclei. Cluster 
configurations in light nuclei can also be  described \cite{Zhang}
in terms of RHO 
symmetries, which underlie the geometrical structure of the 
Bloch--Brink $\alpha$-cluster model \cite{Brink}. 
The 3-dim RHO is also of interest for the interpretation 
of the observed shell structure in atomic clusters
\cite{deHeer,Brack}, especially 
after the 
realization that large deformations can occur in such systems \cite{Bulgac}. 
(See section 9 for further discussion of atomic clusters.) 

The two-dimensional  and
three-dimensional \cite{Barut}  anisotropic harmonic
oscillators have been the subject of several investigations, both at the
classical and the quantum mechanical level (see 
\cite{IJMPA,ht218} for references). 
These oscillators are examples
of superintegrable systems. The special cases with frequency
ratios 1:2 and 1:3 have also been considered \cite{PRAR3407}. While
at the classical level it is clear that the su(N) or sp(2N,R) algebras can
be used for the description of the N-dimensional anisotropic oscillator, the
situation at the quantum level, even in the two-dimensional case, is not as
simple.
It has been proved that a generalized deformed u(2)
algebra is the symmetry algebra of the two-dimensional anisotropic quantum
harmonic oscillator \cite{IJMPA}, which is the oscillator describing the 
single-particle
level spectrum of ``pancake'' nuclei, i.e. of triaxially deformed nuclei
with $\omega_x >> \omega_y$, $\omega_z$. Furthermore, a generalized 
deformed u(3) algebra turns out to be the symmetry algebra of the 
three-dimensional RHO \cite{ht218}. 

\bigskip
{\bf 7. Three-dimensional $q$-deformed (isotropic) harmonic oscillator}
\medskip

Recently the 3-dimensional $q$-deformed (isotropic) harmonic oscillator
has been studied in detail \cite{Terziev}, 
following the mathematical developments of \cite{Jeugt,RRTBI,RRITer}. 
It turns out that in this framework, one can reproduce 
level schemes similar to the ones occuring in the modified harmonic 
oscillator model, first suggested by Nilsson \cite{Nilsson1,Nilsson2}. 
An appropriate $q$-deformed spin--orbit interaction term has also been
developed \cite{Terziev}. Including this term in the 3-dimensional 
$q$-deformed (isotropic) harmonic oscillator scheme one can reproduce 
level schemes similar to these provided by the modified harmonic 
oscillator with spin--orbit interaction. 
It is expected 
that this scheme, without the spin--orbit interaction term, 
will be appropriate for describing the magic numbers occuring in the 
various kinds of atomic clusters \cite{deHeer,Brack}, since 
a description of magic 
numbers of atomic clusters in terms of a Nilsson model without a
spin--orbit interaction has already been attempted \cite{Clem}. 
This subject will be discussed in some detail in Section 9. 

A recent review of the applications of quantum algebraic techniques to
nuclear structure problems can be found in \cite{Roman}. 

\bigskip
{\bf 8. The use of quantum algebras in molecular structure}
\medskip

Similar techniques can be applied
in describing properties of diatomic and polytomic molecules. A brief
list will be given here. 

1) Rotational spectra of diatomic molecules have been described in terms of 
the su$_q$(2) model \cite{CPL175}. 
 As in the case of nuclei, $q$ is a phase factor 
($q=e^{i\tau}$). In molecules $\tau$ is of the order of 0.01. 
The use of the su$_q$(2) symmetry leads to a partial summation of the Dunham
expansion describing the rotational--vibrational spectra of diatomic 
molecules \cite{CPL175}. 
Molecular backbending (bandcrossing) has also been
described in this framework \cite{MRM}. 
Rotational spectra of symmetric 
top molecules have also been considered \cite{Chang1400,Kundu} 
in the framework of the su$_q$(2) symmetry. 
Furthermore, two $q$-deformed rotators with slightly different parameter 
values have been used \cite{RMD} for the description of $\Delta I=1$ staggering
effects in rotational bands of diatomic molecules. (For a discussion of 
$\Delta I=2$ staggering effects in diatomic molecules see \cite{PRAR2533}). 

2) Vibrational spectra of diatomic molecules have been described in terms of 
$q$-deformed anharmonic oscillators having the su$_q$(1,1) \cite{JPAL403}
or the u$_q$(2) $\supset$ o$_q$(2) \cite{CPL178}
symmetry, as well as in terms of generalized deformed oscillators
similar to the ones described in sec. 3 \cite{PRA75,CGY}. 
These results, combined with 1), lead 
to the full summation of the Dunham expansion \cite{JPAL403,CPL178}. 
A two-parameter deformed anharmonic oscillator with u$_{qp}$(2) $\supset$ 
o$_{qp}$(2) symmetry has also been considered \cite{ZZH}. 

3) The physical content of the anharmonic oscillators mentioned in 2) 
has been clarified by constructing WKB equivalent potentials (WKB-EPs) 
and classical equivalent potentials \cite{BDKJMP}
providing approximately the same 
spectrum. 
The results have been corroborated by the study of the 
relation between su$_q$(1,1) and the anharmonic oscillator with  $x^4$ 
anharminicities \cite{Narg}. Furthermore 
the WKB-EP corresponding to the su$_q$(1,1) anharmonic 
oscillator has been connected to a class of Quasi-Exactly Soluble Potentials 
(QESPs) \cite{PLA199}.                                                         

4) Generalized deformed oscillators 
giving the same spectrum as the Morse potential 
\cite{CPL203} and the modified 
P\"oschl--Teller potential \cite{Das2261},  
as well as a deformed oscillator
containing them as special cases \cite{Jann} 
have also been constructed. 
In addition,  $q$-deformed versions of the Morse potential have been given, 
either by using the so$_q$(2,1) symmetry \cite{Cooper}
or by solving a 
$q$-deformed Schr\"odinger equation for the usual Morse potential 
\cite{Dayiht015}. 
For the sake of completeness it should be mentioned that 
a deformed oscillator giving the same spectrum as the Coulomb 
potential has also been constructed \cite{Das4157}.  

5) A $q$-deformed version of the vibron model for diatomic molecules has been 
constructed \cite{PRA1088}, 
in a way similar to that described in sec. 5.  

6) For vibrational spectra of polyatomic molecules a model of $n$ coupled 
generalized deformed oscillators has been built \cite{PRA3611}, 
containing the 
approach of Iachello and Oss \cite{Oss} 
as a special case. In addition
a model of two $Q$-deformed oscillators coupled so that the total 
Hamiltonian has the su$_Q$(2) symmetry has been proved \cite{JCP605}
to be 
equivalent, to lowest order approximation, to a system of two identical Morse
oscillators coupled by the cross-anharmonicity usually used 
empirically in describing vibrational spectra of diatomic molecules.  

7) Quasi-molecular resonances in the systems $^{12}$C+$^{12}$C and 
$^{12}$C+$^{16}$O have been described in terms of a $q$-deformed oscillator
plus a rigid rotator \cite{CY325}.

A review of several of the above topics, concerning the applications of 
quantum algebraic techniques to molecular structure,
accompanied by a detailed and 
self-contained introduction to quantum algebras, has been given by Raychev
\cite{Ray239}. 

\bigskip
{\bf 9. The 3-dimensional $q$-deformed harmonic oscillator and magic 
numbers of alkali metal clusters}
\medskip

Metal clusters have been recently the subject of many investigations
(see \cite{deHeer,Brack,Nester} for relevant reviews). One of the first 
fascinating findings 
in their study was the appearance of magic numbers 
\cite{Martin,Bjorn,Knight1,Knight2,Peder,Brec,Persson}, analogous to 
but different from the magic numbers appearing in the shell structure of 
atomic nuclei \cite{Mayer}. 
This analogy led to the early description of metal 
clusters in terms of the Nilsson--Clemenger model \cite{Clem},
which is a simplified version of the Nilsson model \cite{Nilsson1,Nilsson2} 
of atomic 
nuclei, in which no spin-orbit interaction is included. Further theoretical
investigations in terms of the jellium model \cite{Ekardt,Beck} 
demonstrated that the mean field potential in the case of simple metal 
clusters bears great similarities to the Woods--Saxon potential 
of atomic nuclei, with a slight modification of the ``wine bottle''
type \cite{Kotsos}. 
The Woods--Saxon potential itself looks like a harmonic 
oscillator truncated at a certain energy value and flattened at the bottom. 
It should also be recalled that an early schematic explanation of the 
magic numbers of metallic clusters has been given in terms of a scheme 
intermediate between the level scheme of the 3-dimensional harmonic 
oscillator and the square well \cite{deHeer}. Again in this case the 
intermediate 
potential resembles a harmonic oscillator flattened at the bottom.  

On the other hand, modified versions of harmonic oscillators 
\cite{Bie,Mac} have been recently investigated, as it has already been
mentioned. The spectra of $q$-deformed oscillators increase either 
less rapidly (for $q$ being a phase factor, i.e. $q=e^{i\tau}$ with 
$\tau$ being real) or more rapidly (for $q$ being real, i.e. $q=e^{\tau}$ 
with $\tau$ being real) in comparison to the equidistant spectrum 
of the usual harmonic oscillator \cite{Roman}, while the corresponding 
(WKB-equivalent) potentials \cite{BDKJMP}
resemble the harmonic oscillator potential,
truncated at a certain energy (for $q$ being a phase factor) 
or not (for $q$ being real), 
the deformation inflicting an overall
widening or narrowing of the potential, depending on the value of the 
deformation parameter $q$.   

Very recently, a $q$-deformed version of the 3-dimensional harmonic 
oscillator has been constructed \cite{Terziev}, taking advantage of the 
u$_q$(3) $\supset$ so$_q$(3) symmetry \cite{Smirnov,Jeugt}. 
As it has already been mentioned in Section 7, 
the spectrum of this 3-dimensional $q$-deformed harmonic oscillator 
has been found \cite{Terziev} to reproduce very well the spectrum of the 
modified harmonic oscillator introduced by Nilsson 
\cite{Nilsson1,Nilsson2}, without the 
spin-orbit interaction term. Since the Nilsson model without the 
spin orbit term is essentially the Nilsson--Clemenger model used 
for the description of metallic clusters \cite{Clem}, it is worth examining 
if the 3-dimensional $q$-deformed harmonic oscillator can reproduce 
the magic numbers of simple metallic clusters. This is the subject 
of the present section. 

The space of the 3-dimensional $q$-deformed harmonic oscillator consists of 
the completely symmetric irreducible representations of the quantum algebra
u$_q$(3). In this space a deformed angular momentum algebra, so$_q$(3), 
can be defined \cite{Terziev}. 
The Hamiltonian of the 3-dimensional $q$-deformed 
harmonic oscillator is defined so that it satisfies the following 
requirements:

a) It is an so$_q$(3) scalar, i.e. the energy is simultaneously measurable
with the $q$-deformed  angular momentum related to the algebra so$_q$(3) 
and its $z$-projection.   

b) It conserves the number of bosons, in terms of which the quantum 
algebras u$_q$(3) and so$_q$(3) are realized. 

c) In the limit $q\to 1$ it is in agreement with the Hamiltonian of the usual 
3-dimensional harmonic oscillator. 
 
It has been proved \cite{Terziev} that the Hamiltonian of the 3-dimensional 
$q$-deformed harmonic oscillator satisfying the above requirements 
takes the form
\begin{equation}
H_q = \hbar \omega_0 \left\{ [N] q^{N+1} - {q(q-q^{-1})\over [2] } C_q^{(2)}
\right\},
\end{equation}
where $N$ is the number operator and $C_q^{(2)}$ is the second order 
Casimir operator of the algebra so$_q$(3), while 
\begin{equation}
[x]= {q^x-q^{-x} \over q-q^{-1}}
\end{equation} 
is the definition of $q$-numbers and $q$-operators. 

The energy eigenvalues of the 3-dimensional $q$-deformed harmonic oscillator 
are then \cite{Terziev}
\begin{equation}
E_q(n,l)= \hbar \omega_0 \left\{ [n] q^{n+1} - {q(q-q^{-1}) \over [2]}
[l] [l+1] \right\}, 
\end{equation}
where $n$ is the number of vibrational quanta and $l$ is the eigenvalue of the 
angular momentum, obtaining the values
$l=n, n-2, \ldots, 0$ or 1.  

In the limit of $q\to 1$ one obtains ${\rm lim}_{q\to 1} E_q(n,l)=
\hbar \omega_0 n$, which coincides with the classical result. 

For small values of the deformation parameter $\tau$ (where $q=e^{\tau}$)
one can expand eq. (3) in powers of $\tau$  obtaining \cite{Terziev}
$$
E_q(n,l)= \hbar \omega_0 n -\hbar \omega_0 \tau (l(l+1)-n(n+1))
$$
\begin{equation}
-\hbar \omega_0 \tau^2 \left( l(l+1)-{1\over 3} n(n+1)(2n+1) \right)
+ {\cal O} (\tau^3).
\end{equation}

The last expression to leading order bears great similarity to the modified 
harmonic 
oscillator suggested by Nilsson \cite{Nilsson1,Nilsson2} 
(with the spin-orbit term omitted)
\begin{equation}
V= {1 \over 2} \hbar \omega \rho^2 -\hbar \omega \kappa' 
({\bf L}^2 - <{\bf L}^2>_N ), \qquad \rho=r \sqrt {M\omega \over \hbar} ,
\end{equation}
where 
\begin{equation}
<{\bf L}^2>_N = {N(N+3)\over 2}.
\end{equation}
It has been proved \cite{Terziev} that the spectrum of the 3-dimensional 
$q$-deformed harmonic oscillator closely reproduces the spectrum of 
the modified harmonic oscillator of Nilsson. In both cases the effect 
of the $l(l+1)$ term is to flatten the bottom of the harmonic 
oscillator potential, thus making it to resemble the Woods--Saxon 
potential. 

The level scheme of the 3-dimensional $q$-deformed harmonic oscillator 
(for $\hbar \omega_0 =1$ and $\tau = 0.038$) is given in Table 1
\cite{KarTer}, up to 
a certain energy. Each level is characterized by the quantum numbers 
$n$ (number of vibrational quanta) and $l$ (angular momentum). Next 
to each level its energy, the number of particles it can accommodate
(which is equal to $2(2l+1)$) and the total number of particles up to 
and including this level are given. If the energy difference between 
two successive levels is larger than 0.39, it is  considered as a gap 
separating two successive shells and the energy difference is reported 
between the two levels. In this way magic numbers can be easily read 
in the table: they are the numbers appearing above the gaps, written in 
boldface characters. 

The magic numbers provided by the 3-dimensional $q$-deformed harmonic 
oscillator in Table 1 are compared to available experimental data for 
Na clusters \cite{Martin,Bjorn,Knight1,Peder,Brec}
in Table 2 (columns 2--6) \cite{KarTer}. The following comments apply:

i) Only magic numbers up to 1500
are reported, since it is known that filling of electronic shells 
is expected to occur only up to this limit \cite{Martin}. For large 
clusters beyond this point it is known that magic numbers can be explained by
the completion of icosahedral or cuboctahedral shells of atoms \cite{Martin}. 

ii) Up to 600 particles there is consistency among the various experiments 
and between the experimental results in one hand and our findings in the 
other. 

iii) Beyond 600 particles the predictions of the three  experiments,
which report magic numbers in this region, are 
quite different. However, the predictions of all three  experiments are 
well accommodated by the present model. Magic numbers 694, 832, 1012
are supported by the findings of both Martin {\it et al.} \cite{Martin}
and Br\'echignac {\it et al.} \cite{Brec}, magic numbers 
1206, 1410 are in agreement with the experimental findings of Martin 
{\it et al.} \cite{Martin}, magic numbers 912, 1284 are supported by 
the findings of Br\'echignac {\it et al.}, 
while magic numbers 676, 1100, 1314, 1502 
are in agreement with the experimental findings of Pedersen {\it et al.}
\cite{Peder}. 

In Table 2 the predictions of three simple theoretical models \cite{Mayer}
(non-deformed 3-dimensional harmonic oscillator (column 9), 
square well potential  
(column 8), rounded square well potential (intermediate between the 
previous two, column 7)~) are also reported for comparison. It is clear 
that the predictions of the non-deformed 3-dimensional harmonic oscillator are 
in agreement with the experimental data only up to magic number 40, 
while the other two models give correctly a few more magic numbers (58, 
92, 138), although they already fail by predicting magic numbers at 68, 70, 
106, 112, 156, which are not observed.  

It should be noticed at this point that the first few magic numbers of 
alkali clusters (up to 92) can be correctly reproduced by the assumption 
of the formation of shells of atoms instead of shells of delocalized 
electrons \cite{Anagnos}, this assumption being applicable  under conditions 
not favoring delocalization of the valence electrons of alkali atoms. 

Comparisons among the present results, experimental data 
(by Martin {\it et al.} \cite{Martin} (column 2), Pedersen {\it et al.}
\cite {Peder} (column 3) and Br\'echignac {\it et al.} \cite{Brec}
(column 4)~) and 
theoretical predictions more sophisticated than these reported in Table 2,
can be made in Table 3 \cite{KarTer}, where magic numbers predicted by various 
jellium model calculations (columns 5--8, 
\cite{Martin,Bjorn,Brack,Bulgac}), Woods--Saxon 
and wine bottle potentials (column 9, \cite{Nishi}), as well as by a 
classification scheme using the $3n+l$ pseudo quantum number 
(column 10, \cite{Martin}) are reported. The following observations can be 
made:

i) All magic numbers predicted by the 3-dimensional $q$-deformed harmonic 
oscillator are supported by at least one experiment, with no exception.

ii) Some of the jellium models, as well as the $3n+l$ classification scheme, 
predict magic numbers at 186, 540/542, which are not supported by 
experiment. Some jellium models also predict a magic number at 
748 or 758, again without support from experiment. The Woods--Saxon 
and wine bottle potentials of Ref. \cite{Nishi} predict a magic number at 
68, for which no experimental support exists. The present scheme 
avoids problems at these numbers. It should be noticed, however, 
that in the cases of 186 and 542 the energy gap following them 
in the present scheme is 
0.329 and 0.325 respectively (see Table 1), i.e. quite close to 
the threshold of 0.39 which we have considered as the minimum energy 
gap separating different shells. One could therefore qualitatively 
remark that 186 and 542 are ``built in'' the present scheme as
``secondary'' (not very pronounced) magic numbers.  

The following general remarks can also be made:

i) It is quite remarkable that the 3-dimensional $q$-deformed harmonic 
oscillator reproduces the magic numbers at least as accurately as other,
more sophisticated, models by using only one free parameter ($q=e^{\tau}$). 
Once the parameter is fixed, the whole spectrum is fixed and no further 
manipulations can be made.
This can be considered as evidence that the 3-dimensional $q$-deformed 
harmonic oscillator owns a symmetry (the u$_q$(3) $\supset$ so$_q$(3)
symmetry) appropriate for the description of the physical systems under 
study. 

ii) It has been remarked \cite{Martin} that if $n$ is the number of nodes 
in the solution of the radial Schr\"odinger equation and $l$ is the 
angular momentum quantum number, then the degeneracy of energy levels of 
the hydrogen atom characterized by the same $n+l$ is due to the so(4) 
symmetry of this system, while the degeneracy of energy levels of the 
spherical harmonic oscillator (i.e. of the 3-dimensional isotropic 
harmonic oscillator) characterized by the same $2n+l$ 
is due to the su(3) symmetry of this system. $3n+l$ has been used 
\cite{Martin} to approximate the magic numbers of alkali metal clusters
with some success, but no relevant Lie symmetry could be determined. In view
of the present findings the lack of Lie symmetry related to $3n+l$ is quite 
clear: the symmetry of the system appears to be a quantum algebraic 
symmetry (u$_q$(3)), which is a nonlinear extension of the Lie 
symmetry u(3). 

iii) An interesting problem is to determine a WKB-equivalent potential 
giving (within this approximation) the same spectrum as the 
3-dimensional $q$-deformed harmonic oscillator, using methods similar 
to these  of Ref. \cite{BDKJMP}. The similarity
between the results of the present model and these provided by the 
Woods--Saxon potential (column 9 in Table 3) suggests that the answer 
should be a harmonic oscillator potential flattened at the bottom, 
similar to the Woods--Saxon potential. If such a WKB-equivalent 
potential will show any similarity to a wine bottle shape,
as several potentials used for the description of metal clusters do
\cite{Ekardt,Beck,Kotsos},  remains to be seen. 

In summary, we have shown in this section 
that the 3-dimensional $q$-deformed harmonic 
oscillator with u$_q$(3) $\supset$ so$_q$(3) symmetry correctly 
predicts all experimentally observed magic numbers of alkali metal clusters  
up to 1500, which is the expected limit of validity for theories based on 
the filling of electronic shells. This indicates that u$_q$(3), which 
is a nonlinear deformation of the u(3) symmetry of the spherical
(3-dimensional isotropic) harmonic oscillator, is a good candidate for 
being the symmetry of systems of alkali metal clusters.  

This work  has been supported by the Greek Secretariat 
of Research and Technology under contract PENED 95/1981.

%\bigskip
%\centerline{\bf References}
%\medskip
% \parindent=0pt

\parindent=0pt
%%%%%%%%%%%%%%%%%%%%%%%%%%%%%%%%%%%%%%%%%%%%%%%%%%%%%%%%%%%%%%%%%%%%%%
%%%%%%%%%%%%%%%%%%% Table 1 %%%%%%%%%%%%%%%%%%%%%%%%%%%%%%%%%%%%%%%%

\begin{table}
\caption{
Energy spectrum, $E_q(n,l)$,  of the 3-dimensional $q$-deformed 
harmonic oscillator (eq. (3)), for $\hbar \omega_0 =1$ and 
$q=e^\tau$ with $\tau = 0.038$. Each level is characterized by $n$ 
(the number of vibrational quanta) and  $l$ (the angular momentum).
$2(2l+1)$ represents the number of particles each level can accommodate,
while under ``total'' the total number of particles up to and including 
this level is given. Magic numbers, reported in boldface, correspond to 
energy gaps larger than 0.39, reported between the relevant couples of 
energy levels. }

\bigskip
\newpage

\begin{tabular}{r r r r r || r r r r r}
\hline
$n$ & $l$ & $E_q(n,l)$ & $2(2l+1)$ & total & $n$ & $l$ & $E_q(n,l)$ & 
$2(2l+1)$ & total \\
\hline
 0&  0&  0.000 &  2  &  {\bf 2}  &    9&  5&  12.215 & 22&  462 \\
  &   &  1.000 &     &           &   11& 11&  12.315 & 46&  508 \\
 1&  1&  1.000 &  6  &  {\bf 8}  &   10&  8&  12.614 & 34&  542 \\
  &   &  1.006 &     &           &    9&  3&  12.939 & 14&  {\bf 556} \\
 2&  2&  2.006 & 10  & 18  &     &   &   0.397 &   &      \\
 2&  0&  2.243 &  2  & {\bf 20}  &    9&  1&  13.336 &  6&  562 \\
  &   &  0.780 &     &           &   12& 12&  13.721 & 50&  612 \\
 3&  3&  3.023 & 14  & {\bf 34}  &   10&  6&  13.863 & 26&  638 \\
  &   &  0.397 &     &           &   11&  9&  14.154 & 38&  {\bf 676} \\
 3&  1&  3.420 &  6  & {\bf 40}  &     &   &   0.603 &   &      \\
  &   &  0.638 &     &           &   10&  4&  14.757 & 18&  {\bf 694} \\
 4&  4&  4.058 & 18  & {\bf 58}  &     &   &   0.449 &   &      \\
  &   &  0.559 &     &           &   13& 13&  15.206 & 54&  748 \\
 4&  2&  4.617 & 10  & 68        &   10&  2&  15.316 & 10&  758 \\
 4&  0&  4.854 &  2  & 70        &   10&  0&  15.554 &  2&  760 \\
 5&  5&  5.116 & 22  & {\bf 92}  &   11&  7&  15.592 & 30&  790 \\
  &   &  0.724 &     &           &   12& 10&  15.777 & 42&  {\bf 832} \\
 5&  3&  5.841 & 14  &106        &     &   &   0.884 &   &      \\
 6&  6&  6.204 & 26  &132        &   11&  5&  16.660 & 22&  854 \\
 5&  1&  6.238 &  6  &{\bf 138}  &   14& 14&  16.779 & 58&  {\bf 912} \\
  &   &  0.860 &     &           &     &   &   0.606 &   &      \\
 6&  4&  7.098 & 18  &156        &   11&  3&  17.385 & 14&  926 \\
 7&  7&  7.328 & 30  &186        &   12&  8&  17.410 & 34&  960 \\
 6&  2&  7.657 & 10  &196        &   13& 11&  17.490 & 46& 1006 \\
 6&  0&  7.895 &  2  &{\bf 198}  &   11&  1&  17.782 &  6& {\bf 1012} \\
  &   &  0.502 &     &           &     &   &   0.667 &   &      \\
 7&  5&  8.396 & 22  &220        &   15& 15&  18.449 & 62& 1074 \\
 8&  8&  8.494 & 34  &{\bf 254}  &   12&  6&  18.660 & 26& {\bf 1100} \\
  &   &  0.627 &     &           &     &   &   0.645 &   &      \\
 7&  3&  9.121 & 14  &{\bf 268}  &   14& 12&  19.305 & 50& 1150 \\
  &   &  0.397 &     &           &   13&  9&  19.330 & 38& 1188 \\
 7&  1&  9.518 &  6  &274        &   12&  4&  19.554 & 18& {\bf 1206} \\
 9&  9&  9.709 & 38  &312        &     &   &   0.559 &   &      \\
 8&  6&  9.743 & 26  &{\bf 338}  &   12&  2&  20.113 & 10& 1216 \\
  &   &  0.894 &     &           &   16& 16&  20.226 & 66& 1282 \\
 8&  4& 10.637 & 18  &356        &   12&  0&  20.350 &  2& {\bf 1284} \\
10& 10& 10.980 & 42  &398        &     &   &   0.417 &   &      \\
 9&  7& 11.146 & 30  &428        &   13&  7&  20.767 & 30& {\bf 1314} \\
 8&  2& 11.196 & 10  &438        &     &   &   0.464 &   &      \\
 8&  0& 11.434 &  2  &{\bf 440}  &   15& 13&  21.231 & 54& 1368 \\
  &   &  0.781 &     &           &   14& 10&  21.360 & 42& {\bf 1410} \\
\hline
\end{tabular}
\end{table}

\newpage

%%%%%%%%%%%%%%%%%%% Table 2 %%%%%%%%%%%%%%%%%%%%%%%%%%%%%%%%%%%%%%%%

\begin{table}

\caption{ Magic numbers provided by the 3-dimensional $q$-deformed harmonic 
oscillator (Table 1), reported in column 1, are compared to the experimental 
data of Martin {\it et al.} [111] (column 2), 
Bj{\o}rnholm {\it et al.} [112] (column 3), 
Knight {\it et al.} [113] (column 4), 
Pedersen {\it et al.} [115] (column 5) and 
Br\'echignac {\it et al.} [116] (column 6), concerning Na clusters.
The magic numbers provided [118] by the (non-deformed) 3-dimensional harmonic 
oscillator (column 9), the square well potential (column 8) and a rounded
square well potential intermediate between the previous two (column 7) 
are also shown for comparison. See text for discussion. }

\bigskip

\begin{tabular}{c c c c c c c c c}
\hline
    &   exp.  & exp.    &   exp.   &  exp.& exp.  & int. & sq. well & h. osc.\\
%  our    Martin     Bjorn        Knight  Peder   Brec int    well        HO
 present & [111] & [112]& [113]& [115]& [116] & [118] & [118] & [118] \\  
\hline
    2  &    2      &   2     & 2&      &      &     2   &   2     &   2 \\
    8  &    8      &   8     & 8&      &      &     8   &   8     &   8 \\
  (18) &   18      &         &  &      &      &    18   &  18     &     \\
   20  &   20      &  20     &20&      &      &    20   &  20     &  20 \\
   34  &   34      &         &  &      &      &    34   &  34     &     \\
   40  &   40      &  40     &40&   40 &      &    40   &  40     &  40 \\
   58  &   58      &  58     &58&   58 &      &    58   &  58     &     \\
       &           &         &  &      &      &  68,70  &  68     &  70 \\
   92  &  90,92    &  92     &92&   92 &   93 &    92   & 90,92   &     \\
       &           &         &  &      &      &  106,112&  106    & 112 \\
  138  &  138      & 138     &  &  138 &  134 &    138  &  132,138&     \\
  198  &  198$\pm$2 & 196     & &  198 &  191 &    156  &   156   & 168 \\
  254  &           & 260$\pm$4& &      &      &         &         &     \\
  268  &  263$\pm$5 &         & &  264 &  262 &         &         &     \\
  338  & 341$\pm$5 & 344$\pm$4& &  344 &  342 &         &         &     \\
  440  & 443$\pm$5 & 440$\pm$2& &  442 &  442 &         &         &     \\
  556  & 557$\pm$5 & 558$\pm$8& &  554 &  552 &         &         &     \\
  676  &           &         &  &  680 &      &         &         &     \\
  694  &  700$\pm$15&         & &      &  695 &         &         &     \\
  832  &  840$\pm$15&         & &  800 &  822 &         &         &     \\
  912  &           &         &  &      &  902 &         &         &     \\
 1012  & 1040$\pm$20&         & &  970 & 1025 &         &         &     \\
 1100  &           &         &  & 1120 &      &         &         &     \\
 1206  & 1220$\pm$20&         & &      &      &         &         &     \\
 1284  &           &         &  &      & 1297 &         &         &     \\
 1314  &           &         &  & 1310 &      &         &         &     \\
 1410  &  1430$\pm$20&        & &      &      &         &         &     \\
 1502  &           &         &  & 1500 &      &         &         &     \\
\hline
\end{tabular}
\end{table}

\newpage
%%%%%%%%%%%%%%%%%%% Table 3 %%%%%%%%%%%%%%%%%%%%%%%%%%%%%%%%%%%%%%%%

\begin{table}

\caption{Magic numbers provided by the 3-dimensional $q$-deformed harmonic 
oscillator (Table 1), reported in colunmn 1, are compared to the experimental 
data of Martin {\it et al.} [111] (column 2),
Pedersen {\it et al.} [115] (column 3), and 
Br\'echignac {\it et al.} [116] (column 4),
as well as to the theoretical predictions of various jellium model 
calculations reported by Martin {\it et al.} [111] (column 5), Bj{\o}rnholm 
{\it et al.} [112] (column 6), Brack [73] (column 7), Bulgac and Lewenkopf
[74] (column 8), 
the theoretical predictions of 
Woods--Saxon and wine bottle potentials reported by Nishioka {\it et al.}
[125] (column 9),
as well as to the magic numbers predicted by the  classification scheme 
using the $3n+l$ pseudo quantum number, reported by Martin {\it et al.}
[111] (column 10). See text for discussion.  
}

\bigskip

\begin{tabular}{c c c c c c c c c c}
\hline
         &exp. &exp. &exp. &jell.   & jell. &  jell.  & jell. & WS   &$3n+l$ \\
 present & [111] & [115] &  [116] & [111] & [112] & [73] 
& [74]& [125]  & [111] \\  
\hline
%          exp   exp   exp jell        jell     jell          
%  our    Martin Ped  Brec Martin      Bjorn    Brack Bulgac  Nishi    3n+l 
  2 &    2     &     &     &  2     &    2 &     2   &       &     2  &    2 \\
  8 &    8     &     &     &  8     &    8 &     8   &       &     8  &    8 \\
(18)&   18     &     &     & 18     &   18 &         &       &        &   18 \\
 20 &   20     &     &     &(20)    &   20 &    20   &       &    20  &      \\
 34 &   34     &     &     & 34     &   34 &    34   &  34   &        &   34 \\
 40 &   40     &  40 &     &(40)    &   40 &         &       &    40  &      \\
 58 &   58     &  58 &     & 58     &   58 &    58   &  58   &    58  &   58 \\
    &          &     &     &        &      &         &       &    68  &      \\
 92 &  90,92   &  92 &  93 & 92     &   92 &    92   &  92   &    92  &   90 \\
138 &  138     & 138 & 134 &134     &  138 &   138   &  138  &   138  &  132 \\
    &          &     &     &186     &  186 &   186   &  186  &        &  186 \\
198 & 198$\pm$2& 198 & 191 &(196)   &  196 &         &       &   198  &      \\
254 &          &     &     &254     &  254 &   254   &  254  &   254  &  252 \\
268 & 263$\pm$5& 264 & 262 &(268)   &      &         &       &   268  &      \\
338 & 341$\pm$5& 344 & 342 &338(356)&  338 &   338   &  338  &   338  &  332 \\
440 & 443$\pm$5& 442 & 442 &440     &  440 & 438,440 &  440  &   440  &  428 \\
    &          &     &     &        &      &   542   &  542  &        &  540 \\
556 & 557$\pm$5& 554 & 552 &562     &  556 &   556   &  556  &   562  &      \\
676 &          & 680 &     &        &  676 &   676   &  676  &        &  670 \\
694 &700$\pm$15&     & 695 &704     &      &         &       &   694  &      \\
    &          &     &     &        &      &   758   &  748  &        &      \\
832 &840$\pm$15& 800 & 822 &852     &  832 &   832   &  832  &   832  &  820 \\
912 &          &     & 902 &        &      &   912   &  912  &        &      \\
1012&1040$\pm$20&970 &1025 &        &      &  1074   & 1074  &  1012  &  990 \\
1100&          &1120 &     &        &      &  1100   & 1100  &  1100  &      \\
1206&1220$\pm$20&    &     &        &      &         &       &  1216  & 1182 \\
1284&          &     &1297 &        &      &  1284   & 1284  &        &      \\
1314&          &1310 &     &        &      &         &       &  1314  &      \\
1410&1430$\pm$20&    &     &        &      &         &       &        & 1398 \\
1502&          &1500 &     &        &      &  1502   &  1502 &  1516  &      \\
 \hline
\end{tabular}
\end{table}


\newpage
\begin{thebibliography}{99}

%%%%%%%%%%%%%%%%%%%% introduction %%%%%%%%%%%%%%%%%%%%%%%%%%

\bibitem{Chari}
V. Chari and A. Pressley, {\it A Guide to Quantum Groups} (Cambridge 
University Press, Cambridge, 1994). 

\bibitem{BieL}
L. C. Biedenharn and M. A. Lohe, {\it Quantum Group Symmetry and $q$-Tensor
Algebras} (World Scientific, Singapore, 1995). 

\bibitem{KlimykS}
A. Klimyk and K. Schm\"udgen, {\it Quantum Groups and their Representations}
(Springer, Berlin, 1997). 

\bibitem{Bie}  
L. C.  Biedenharn, {\it J. Phys. A} {\bf 22} (1989) L873.

\bibitem{Mac}   
A. J. Macfarlane, {\it J. Phys. A} {\bf 22} (1989) 4581.

\bibitem{ArikCoon}
 M.  Arik and D. D. Coon, {\it J. Math. Phys.} {\bf 17} (1976) 524.

\bibitem{McD}
R. J. McDermott and A. I. Solomon, {\it J. Phys. A} {\bf 27} (1994) L15. 

\bibitem{KiblerN} 
M. Kibler and T. N\'egadi, {\it J. Phys. A} {\bf 24} (1991) 5283. 

\bibitem{ArikAH}
M. Arik, F. Aydin, E. Hizel, J. Kornfilt and A. Yildiz, 
{\it J. Math. Phys.} {\bf 35} (1994) 3074. 

\bibitem{DayiD}
\"O. F. Dayi and I. H. Duru, {\it J. Phys. A} {\bf 28} 
(1995) 2395. 

\bibitem{Cod}
S. Codriansky, {\it Int. J. Theor. Phys.} {\bf 30} (1991) 59. 

\bibitem{ArikCelik}
M. Arik and S. Celik, {\it Z. Phys. C} {\bf 59} (1993) 99. 

\bibitem{ArikUM}
M. Arik, G. \"Unel and M. Mungan, {\it Phys. Lett. B} 
{\bf 321} (1994) 385. 

\bibitem{MMZ}
V. I. Man'ko, G. Marmo and F. Zaccaria, {\it Phys. Lett. A}
{\bf 191} (1994) 13.

\bibitem{MKas}
R. M. Mir-Kasimov, {\it J. Phys. A} {\bf 24} (1991) 4283. 

\bibitem{ArikM} 
M. Arik and M. Mungan, {\it Phys. Lett. B} {\bf 282} (1992) 
101. 

\bibitem{Das}
C. Daskaloyannis, {\it J. Phys. A} {\bf 24} (1991) L789.

\bibitem{ArikDT}
M. Arik, E. Demircan, T. Turgut, L. Ekinci and M. Mungan,  
{\it Z. Phys. C} {\bf 55} (1992) 89.

\bibitem{PLB307} 
D.  Bonatsos and C. Daskaloyannis, {\it Phys. Lett. B}
 {\bf 307} (1993) 100.

\bibitem{PLB331} 
D. Bonatsos, C. Daskaloyannis, D. Ellinas and A. Faessler, 
{\it Phys. Lett. B} {\bf 331} (1994) 150. 

\bibitem{JPA26}
D. Bonatsos, C. Daskaloyannis, and P. Kolokotronis, {\it J.
Phys. A} {\bf 26} (1993) L871; {\it Mod. Phys. Lett. A} {\bf 10} (1995)
2197. 

\bibitem{Pan} 
F. Pan, {\it J. Math. Phys.} {\bf 35} (1994) 5065.

\bibitem{CJP46}
D. Bonatsos, P. Kolokotronis, C. Daskaloyannis, A. Ludu and 
C. Quesne, {\it Czech. J. Phys.} {\bf 46} (1996) 1189. (Presented at 
the 5th International Colloquium on Quantum Groups: {\it Quantum Groups 
and Integrable Systems} (Prague 1996).) 

\bibitem{JMP38}
D. Bonatsos, C. Daskaloyannis, P. Kolokotronis, A. Ludu and 
C. Quesne, {\it J. Math. Phys.} {\bf 38} (1997) 369. 

%%%%%%%%%%%%%%%%%% suq(2) model %%%%%%%%%%%%%%%%%%%%%%%%%%

\bibitem{RRS}
P. P. Raychev, R. P. Roussev and Yu. F. Smirnov, {\it J.
Phys. G} {\bf 16} (1990) L137. 

\bibitem{PLB251}
D. Bonatsos, E. N. Argyres, S. B. Drenska, P. P. Raychev, 
R. P. Roussev and Yu. F. Smirnov, {\it Phys. Lett. B} {\bf 251} (1990) 477. 

\bibitem{JPG17}
D. Bonatsos, S. B. Drenska, P. P. Raychev, R. P. Roussev and 
Yu. F. Smirnov, {\it J. Phys. G} {\bf 17} (1991) L67. 

\bibitem{MRR20}
N. Minkov, R. P. Roussev and P. P. Raychev, {\it J. Phys. 
G} {\bf 20} (1994) L67. 

\bibitem{MRR21}
N. Minkov, P. P. Raychev and R. P. Roussev, {\it J. Phys. G}
{\bf 21} (1995) 557. 

\bibitem{MDRRB}
N. Minkov, S. B. Drenska, P. P. Raychev, R. P. Roussev and 
D. Bonatsos, {\it J. Phys. G} {\bf 22} (1996) 1633. 

\bibitem{JPA3275}
D. Bonatsos, A. Faessler, P. P. Raychev, R. P. Roussev and 
Yu. F. Smirnov, {\it J. Phys. A} {\bf 25} (1992) 3275. 

\bibitem{Muker}
M. Mukerjee, {\it Phys. Lett. B} {\bf 251} (1990) 229. 

\bibitem{LongJi}
G. L. Long and H. Y. Ji, {\it Phys. Rev. C} {\bf 57} (1998) 1686.  

\bibitem{PCWF} 
J. L. Pin, J. Q. Chen, C. L. Wu and D. H. Feng, {\it Phys.
Rev. C} {\bf 43} (1991) 2224. 

\bibitem{ZC} 
N. V. Zamfir and R. F. Casten, {\it Phys. Rev. Lett. } {\bf 75}
(1995) 1280. 

%%%%%%%%%%%%% extensions of suq(2) model %%%%%%%%%%%%%%%%%%%%%%%%%%

\bibitem{PRC50}
D. Bonatsos, C. Daskaloyannis, A. Faessler, P. P. Raychev 
and R. P. Roussev, {\it Phys. Rev. C} {\bf 50} (1994) 497. 

\bibitem{PRC29}
D. Bonatsos and A. Klein, {\it Phys. Rev. C} {\bf 29} (1984) 1879;
{\it At. Data Nucl. Data Tables} {\bf 30} (1984) 27.  

\bibitem{DGGRR52}
S. Drenska, A. Georgieva, V. Gueorguiev, R. Roussev and P. 
Raychev, {\it Phys. Rev. C} {\bf 52} (1995) 1853. 

\bibitem{Holmb}
P. Holmberg and P. O. Lipas, {\it Nucl. Phys. A} {\bf 117}
(1968) 552. 

\bibitem{BMKibler} 
R. Barbier, J. Meyer and M. Kibler, {\it Int. J. Mod. 
Phys. E} {\bf 4} (1995) 385. 

%%%%%%%%%%%%%%%%%%%%%% pairing %%%%%%%%%%%%%%%%%%%%%%%%%%%%%%%%%%

\bibitem{JPAL101}
D. Bonatsos, {\it J. Phys. A} {\bf 25} (1992) L101.

\bibitem{PLB278}
D. Bonatsos and C. Daskaloyannis, {\it Phys. Lett. B}
{\bf 278} (1992) 1.  

\bibitem{JPA1299} 
D. Bonatsos, C. Daskaloyannis and A. Faessler, {\it J. Phys. A}
{\bf 27} (1994) 1299. 

\bibitem{ShaSha} 
S. Shelly Sharma and N. K. Sharma, {\it Phys. Rev. C}
{\bf 50} (1994) 2323. 

\bibitem{AMen}
S. S. Avancini and D. P. Menezes, {\it J. Phys. A} {\bf 26}
(1993) 6261. 

%%%%%%%%%%%%%%  q-deformed nuclear models %%%%%%%%%%%%%%%

\bibitem{JPAL267}
D. Bonatsos, A. Faessler, P. P. Raychev, R. P. Roussev 
and Yu. F. Smirnov, {\it J. Phys. A} {\bf 25} (1992) L267. 

\bibitem{Niigata}
D. Bonatsos, in {\it Symmetries in Science VII: Spectrum 
Generating Algebras and Dynamic Symmetries in Physics (Niigata 1992)}, 
ed. B. Gruber and T. Otsuka (Plenum, New York, 1993) p. 111. 

\bibitem{Cseh25}
J. Cseh, {\it J. Phys. A} {\bf 25} (1992) L1225. 

\bibitem{Gupta20}
R. K. Gupta, {\it J. Phys. G} {\bf 20} (1994) 1067. 

\bibitem{Falco}
L. De Falco, A. Jannussis, R. Mignani and A. Sotiropoulou, 
{\it Mod. Phys. Lett. A} {\bf 9} (1994) 3331. 

\bibitem{GCLGS} 
R. K. Gupta, J. Cseh, A. Ludu, W. Greiner and W. Scheid, 
{\it J. Phys. G} {\bf 18} (1992) L73. 

\bibitem{Mesa}
A. Del Sol Mesa, G. Loyola, M. Moshinsky and V. Vel\'azquez, 
{\it J. Phys. A} {\bf 26} (1993) 1147. 

\bibitem{Cseh19}
J. Cseh, {\it J. Phys. G} {\bf 19} (1993) L63. 

\bibitem{WangYang}
Y. C. Wang and Z. S. Yang, {\it Commun. Theor. Phys.} {\bf 
17} (1992) 449. 

\bibitem{GuptaLudu} 
R. K. Gupta and A. Ludu, {\it Phys. Rev. C} {\bf 48} (1993) 593. 

\bibitem{Pan50}
F. Pan, {\it Phys. Rev. C} {\bf 50} (1994) 1876. 

\bibitem{Jeugt}
J. Van der Jeugt, {\it J. Phys. A} {\bf 25} (1992) L213; {\it J. Math. Phys.}
{\bf 34} (1993) 1799; {\it Can. J. Phys.} {\bf 72} (1994) 519.  

\bibitem{RRTBI}
P. P. Raychev, R. P. Roussev, P. A. Terziev, D. Bonatsos and N. Lo
Iudice, {\it J. Phys. A} {\bf 29} (1996) 6939. 

\bibitem{RRITer}
P. P. Raychev, R. P. Roussev, N. Lo Iudice and P. A. Terziev, {\it J. Phys. 
A} {\bf 30} (1997) 4383. 

\bibitem{MAP}
D. P. Menezes, S. S. Avancini and C. Provid\^encia, 
{\it J. Phys. A} {\bf 25} (1992) 6317. 

\bibitem{FSMen}
F. W. F\'avero, L. O. E. Santos and D. P. Menezes, {\it Int. J. Mod. 
Phys. E} {\bf 4} (1995) 547. 

\bibitem{JPA895}
D. Bonatsos, L. Brito, D. P. Menezes, C. Provid\^encia and 
J. da Provid\^encia, {\it J. Phys. A} {\bf 26} (1993) 895, 5185;
{\it J. Phys. A} {\bf 28} (1995) 1787. 

\bibitem{JPG1209}
C. Provid\^encia, L. Brito, J. da Provid\^encia, D. Bonatsos 
and D. P. Menezes, {\it J. Phys. G} {\bf 20} (1994) 1209; 
{\it J. Phys. G} {\bf 21} (1995) 591.  

\bibitem{PLA192}
D. Bonatsos, S. S. Avancini, D. P. Menezes and C. 
Provid\^encia, {\it Phys. Lett. A} {\bf 192} (1994) 192. 

\bibitem{AEGPL}
S. S. Avancini, A. Eiras, D. Galetti, B. M. Pimentel and 
C. L. Lima, {\it J. Phys. A } {\bf 28} (1995) 4915. 

\bibitem{BPP} 
L. Brito, C. Provid\^encia, J. da Provid\^encia, S. S. 
Avancini, F. F. de Souza Cruz, D. P. Menezes and M. M. Watanabe de Moraes, 
{\it Phys. Rev. A} {\bf 52} (1995) 92. 

%%%%%%%%%%%% anisotropic HO %%%%%%%%%%%%%%%%%%%%%%%%%%%%%%%%5

\bibitem{Nolan}
P. J. Nolan and P. J. Twin, {\it Ann. Rev. Nucl. Part. Sci. } {\bf 38}
(1988) 533.   

\bibitem{Janssens}
R. V. F. Janssens and T. L. Khoo, {\it Ann. Rev. Nucl. Part. Sci. }
{\bf 41} (1991) 321. 

\bibitem{Rae}
W. D. M. Rae, {\it Int. J. Mod. Phys. A} {\bf 3} (1988) 1343. 

\bibitem{Zhang}
J. Zhang, W. D. M. Rae and A. C. Merchant, {\it Nucl. Phys. A} {\bf 575}
(1994) 61. 

\bibitem{Brink}
D. M. Brink, in {\it Proc. Int. School of Physics, Enrico Fermi Course 
XXXVI (Varenna 1966)}, ed. C. Bloch (Academic Press, New York, 1966) p. 247. 

\bibitem{deHeer}
W. A. de Heer, {\it Rev. Mod. Phys.} {\bf 65} (1993) 611. 

\bibitem{Brack}
M. Brack, {\it Rev. Mod. Phys.} {\bf 65} (1993) 677. 

\bibitem{Bulgac}
A. Bulgac and C. Lewenkopf, {\it Phys. Rev. Lett.} {\bf 71} (1993) 4130.  

\bibitem{Barut}
A. O. Barut, {\it Phys. Rev. } {\bf 139} (1965) B1433. 

\bibitem{IJMPA}
D. Bonatsos, P. Kolokotronis, D. Lenis and C. Daskaloyannis, 
{\it Int. J. Mod. Phys. A} {\bf 12} (1997) 3335.  
 
\bibitem{ht218}
D. Bonatsos, C. Daskaloyannis, P. Kolokotronis and D. Lenis, 
hep-th/9411218. 

\bibitem{PRAR3407} 
D.  Bonatsos, C.  Daskaloyannis and K. Kokkotas, {\it Phys. Rev.
A} {\bf 48}  (1993) R3407; {\bf 50} (1994) 3700.

\bibitem{Terziev}
P. P. Raychev, R. P. Roussev, N. Lo Iudice and P. A. Terziev,
{\it J. Phys. G} {\bf 24} (1998) 1. 

\bibitem{Nilsson1}
S. G. Nilsson, {\it Mat. Fys. Medd. Dan. Vid. Selsk. }
{\bf 29} (1955) no 16. 

\bibitem{Nilsson2}
S. G. Nilsson and I. Ragnarsson, {\it Shapes and Shells in Nuclear Structure}
(Cambridge University Press, Cambridge, 1995). 

\bibitem{Clem}
K. Clemenger, Phys. Rev. B 32 (1985) 1359; Ph.D. thesis,
University of California, Berkeley, 1985 (unpublished). 

\bibitem{Roman}
D. Bonatsos, C. Daskaloyannis, P. Kolokotronis and D. Lenis, 
{\it Rom. J. Phys.} {\bf 41} (1996) 109.  

%%%%%%%%%%%%%%%%%% molecules %%%%%%%%%%%%%%%%%%%%%%

\bibitem{CPL175}
D. Bonatsos, P. P. Raychev, R. P. Roussev and Yu. F. Smirnov, {\it Chem.
Phys. Lett.} {\bf 175} (1990) 300. 

\bibitem{MRM}
L. P. Marinova, P. P. Raychev and J. Maruani, {\it Molec. Phys.} {\bf 82}
(1994) 1115. 

\bibitem{Chang1400}
Z. Chang, {\it Phys. Rev. A} {\bf 46} (1992) 1400. 

\bibitem{Kundu}
A. Kundu and Y. J. Ng, {\it Phys. Lett. A} {\bf 197} (1995) 221.  

\bibitem{RMD}
P. Raychev, J. Maruani and S. Drenska, {\it Phys. Rev. A} {\bf 56} (1997) 2759.

\bibitem{PRAR2533}
D. Bonatsos, C. Daskaloyannis, S. B. Drenska, G. A. Lalazissis, N. Minkov, 
P. P. Raychev and  R. P. Roussev, {\it Phys. Rev. A} {\bf 54} (1996) R2533. 

\bibitem{JPAL403}
D. Bonatsos, E. N. Argyres and P. P. Raychev, {\it J. Phys. A} {\bf 24}
(1991) L403. 

\bibitem{CPL178}
D. Bonatsos, P. P. Raychev and A. Faessler, {\it Chem. Phys. Lett.} {\bf 178}
(1991) 221. 

\bibitem{PRA75}
D. Bonatsos and C. Daskaloyannis, {\it Phys. Rev. A} {\bf 46} (1992) 75.  

\bibitem{CGY}
Z. Chang, H. Y. Guo and H. Yan, {\it Commun. Theor. Phys.} {\bf 17} (1992)
183. 

\bibitem{ZZH}
H. Q. Zhou, X. M. Zhang and J. S. He, {\it Mod. Phys. Lett. B} {\bf 9} 
(1995) 1053.  

\bibitem{BDKJMP}
D. Bonatsos, C. Daskaloyannis and K. Kokkotas, {\it J. Phys. A} {\bf 24}
(1991) L795; {\it J. Math. Phys.} {\bf 33}
(1992) 2958; {\it Chem. Phys. Lett.} {\bf 193} (1992) 191;  
{\it Phys. Rev. A} {\bf 45} (1992) R6153.  

\bibitem{Narg}
F. J. Narganes-Quijano, {\it J. Phys. A} {\bf 24} (1991) 1699. 

\bibitem{PLA199}
D. Bonatsos, C. Daskaloyannis and H. A. Mavromatis, {\it Phys. Lett. A} 
{\bf 199} (1995) 1. 

\bibitem{CPL203}
D. Bonatsos and C. Daskaloyannis, {\it Chem. Phys. Lett.} {\bf 203} (1993)
150. 

\bibitem{Das2261}
C. Daskaloyannis, {\it J Phys. A} {\bf 25} (1992) 2261. 

\bibitem{Jann}
A. Jannussis, {\it J. Phys. A} {\bf 26} (1993) L233. 

\bibitem{Cooper}
I. L. Cooper and R. K. Gupta, {\it Phys. Rev. A} {\bf 52} (1995) 941. 

\bibitem{Dayiht015}
\"O. F. Dayi and I. H. Duru, {\it Int. J. Mod. Phys. A} {\bf 12} (1997) 2373. 

\bibitem{Das4157}
C. Daskaloyannis and K. Ypsilantis, {\it J. Phys. A} {\bf 25} (1992) 4157. 

\bibitem{PRA1088}
R. N. Alvarez, D. Bonatsos and Yu. F. Smirnov, {\it Phys. 
Rev. A} {\bf 50} (1994) 1088. 

\bibitem{PRA3611}
D. Bonatsos and C. Daskaloyannis, {\it Phys. Rev. A} {\bf 48} (1993) 3611. 

\bibitem{Oss}
F. Iachello and S. Oss, {\it Chem. Phys. Lett.} {\bf 187} (1991) 500. 

\bibitem{JCP605}
D. Bonatsos, C. Daskaloyannis and P. Kolokotronis, {\it J. Chem. Phys.}
{\bf 106} (1997) 605. 

\bibitem{CY325}
Z. Chang and H. Yan, {\it Commun. Theor. Phys.} {\bf 19} (1993) 325. 

\bibitem{Ray239}
P. P. Raychev, {\it Adv. Quant. Chem.} {\bf 26} (1995) 239. 

%%%%%%%%%%%% atomic clusters %%%%%%%%%%%%%%%%%%%%%%%%%%%%%%%%%%%%%%%%

\bibitem{Nester} 
V. O. Nesterenko, {\it Fiz. Elem. Chastits At. Yadra} {\bf 23} (1992) 1665
[{\it Sov. J. Part. Nucl.} {\bf 23} (1992) 726]. 

\bibitem{Martin}
T. P. Martin, T. Bergmann, H. G\"ohlich and T. lange, {\it Chem. Phys. Lett.}
{\bf 172} (1990) 209; {\it Z. Phys. D} {\bf 19} (1991) 25. 

\bibitem{Bjorn}
S. Bj{\o}rnholm, J. Borggreen, O. Echt, K. Hansen, J. Pedersen and H. D. 
Rasmussen, {\it Phys. Rev. Lett.} {\bf  65} (1990) 1627; 
{\it Z. Phys. D} {\bf  19} (1991) 47.  

\bibitem{Knight1}
W. D. Knight, K. Clemenger, W. A. de Heer, W. A. Saunders, M. Y. Chou 
and M. L. Cohen, {\it Phys. Rev. Lett.} {\bf  52} (1984) 2141. 

\bibitem{Knight2}
W. D. Knight, W. A. de Heer, K. Clemenger and W. A. Saunders, {\it Solid 
State Commun.} {\bf  53} (1985) 445. 

\bibitem{Peder}
J. Pedersen, S. Bj{\o}rnholm, J. Borggreen, K. Hansen, T. P. Martin and H. D. 
Rasmussen, {\it Nature} {\bf 353} (1991) 733. 

\bibitem{Brec}
C. Br\'echignac, Ph. Cahuzac, M. de Frutos, J.-Ph. Roux and K. Bowen, in 
{\it Physics and Chemistry of Finite Systems: From Clusters to Crystals}, 
ed. P. Jena {\it et al.} (Kluwer, Dordrecht, 1992) Vol. 1 p. 369. 

\bibitem{Persson}
J. L. Persson, R. L. Whetten, H. P. Cheng and R. S. Berry, {\it Chem. Phys. 
Lett.} {\bf 186} (1991) 215. 

\bibitem{Mayer}
M. G. Mayer and J. H. D. Jensen, {\it Elementary Theory of Nuclear Shell
Structure} (Wiley, New York, 1955). 

\bibitem{Ekardt}
W. Ekardt, {\it Ber. Bunsenges. Phys. Chem.} {\bf 88} (1984) 289; 
{\it Phys. Rev. B} {\bf 29} (1984) 1558. 

\bibitem{Beck}
D. E. Beck, {\it Solid State Commun.} {\bf 49} (1984) 381; 
{\it Phys. Rev. B} {\bf 30} (1984) 6935.  

\bibitem{Kotsos}
B. A. Kotsos and M. E. Grypeos, in {\it Atomic and Nuclear Clusters},
ed. G. S. Anagnostatos and W. von Oertzen (Springer, Berlin, 1995) p. 242. 

\bibitem{Smirnov}
Yu. F. Smirnov, V. N. Tolstoy and Yu. I. Kharitonov, {\it Yad. Fiz.} {\bf 54}
(1991) 721 [{\it Sov. J. Nucl. Phys.} {\bf 54} (1991) 437];  Yu. F. Smirnov 
and Yu. I. Kharitonov, {\it Yad. Fiz.} {\bf 56} (1993) 263 [{\it Phys. At. 
Nucl.} {\bf  56} (1993) 1143]; {\it Yad. Fiz.} {\bf  58}
(1995) 651 [{\it Phys. At. Nucl.} {\bf  58} (1995) 595]; 
A. A. Malashin, Yu. F. Smirnov and Yu. I. Kharitonov, {\it Yad. Fiz.}
{\bf  58} (1995) 651 [{\it Phys. At. Nucl.} {\bf 58} (1995) 595]; {\it Yad. 
Fiz.} {\bf  58} (1995) 1105 [{\it Phys. At. Nucl.} {\bf  58} (1995) 1031].  

\bibitem{KarTer}
D. Bonatsos, C. Daskaloyannis, N. Karoussos, P. P. Raychev, R. P. Roussev 
and P. A. Terziev, N.C.S.R. ``Demokritos'' and INRNE, Sofia preprint (1998). 

\bibitem{Anagnos}
G. S. Anagnostatos, {\it Phys. Lett. A} {\bf  154} (1991) 169. 

\bibitem{Nishi}
H. Nishioka, K. Hansen and B. R. Mottelson, {\it Phys. Rev. B} {\bf  42}
(1990) 9377. 


\end{thebibliography}
\end{document}